\documentclass[sigconf]{acmart}

\pdfoutput=1
\usepackage{algorithm}
\usepackage{algorithmicx}
\usepackage{algpseudocode} 
\usepackage{graphicx}
\usepackage{float}     
\usepackage{textcomp}
\usepackage{hyperref}
\usepackage{amsfonts}
\usepackage{algpseudocode}

\usepackage{comment} 
\excludecomment{comment} 
\hypersetup{
    colorlinks=true,
    linkcolor=black,
    urlcolor=blue,
    citecolor=black
}
\usepackage{etoolbox}
\makeatletter
\patchcmd{\@makecaption}
  {\scshape}
  {}
  {}
  {}
\makeatother

\AtBeginDocument{%
  \providecommand\BibTeX{{%
    \normalfont B\kern-0.5em{\scshape i\kern-0.25em b}\kern-0.8em\TeX}}}

\setcopyright{acmlicensed}
\copyrightyear{2024}
\acmYear{2024}
\acmDOI{XXXXXXX.XXXXXXX}

\acmConference[arXiv 2024]{arXiv Preprint}{Online}{2024}{}
\acmISBN{978-1-4503-XXXX-X/18/06}

\author{Jing Xu}
\affiliation{
  \institution{Institute of Computing Technology, Chinese Academy of Sciences \ University of Chinese Academy of Sciences}
  \city{Beijing}
  \country{China}
}
\email{xujing@ncic.ac.cn}

\author{Zhan Wang}
\affiliation{
  \institution{Institute of Computing Technology, Chinese Academy of Sciences}
  \city{Beijing}
  \country{China}
}
\email{wangzhan@ncic.ac.cn}

\author{Fan Yang}
\affiliation{
  \institution{Institute of Computing Technology, Chinese Academy of Sciences}
  \city{Beijing}
  \country{China}
}
\email{yangfan@ncic.ac.cn}

\author{Ning Kang}
\affiliation{
  \institution{University of Chinese Academy of Sciences}
  \city{Beijing}
  \country{China}
}
\email{kangning18z@ict.ac.cn}

\author{Zhenlong Ma}
\affiliation{
  \institution{University of Chinese Academy of Sciences}
  \city{Beijing}
  \country{China}
}
\email{mazhenlong@ncic.ac.cn}

\author{Guojun Yuan}
\affiliation{
  \institution{Institute of Computing Technology, Chinese Academy of Sciences}
  \city{Beijing}
  \country{China}
}
\email{yuanguojun@ncic.ac.cn}

\author{Guangming Tan}
\affiliation{
  \institution{Institute of Computing Technology, Chinese Academy of Sciences}
  \city{Beijing}
  \country{China}
}
\email{tgm@ncic.ac.cn}

\author{Ninghui Sun}
\affiliation{
  \institution{Institute of Computing Technology, Chinese Academy of Sciences}
  \city{Beijing}
  \country{China}
}
\email{snh@ict.ac.cn}
\begin{document}
\begin{sloppypar}  

 \renewcommand{\shortauthors}{Jing Xu et al.}


%
\title{FNCC: Fast Notification Congestion Control in Data Center Networks}

\begin{abstract}
Congestion control plays a pivotal role in large-scale data centers, facilitating ultra-low latency, high bandwidth, and optimal utilization. Even with the deployment of data center congestion control mechanisms such as DCQCN and HPCC, these algorithms often respond to congestion sluggishly. This sluggishness is primarily due to the slow notification of congestion. It takes almost one round-trip time (RTT) for the congestion information to reach the sender. In this paper, we introduce the Fast Notification Congestion Control (FNCC) mechanism, which achieves sub-RTT notification. FNCC leverages the acknowledgment packet (ACK) from the return path to carry in-network telemetry (INT) information of the request path, offering the sender more timely and accurate INT. To further accelerate the responsiveness of last-hop congestion control, we propose that the receiver notifies the sender of the number of concurrent congested flows, which can be used to adjust the congested flows to a fair rate quickly. Our experimental results demonstrate that FNCC reduces flow completion time by 27.4\% and 88.9\% compared to HPCC and DCQCN, respectively. Moreover, FNCC triggers minimal pause frames and maintains high utilization even at 400Gbps.

\end{abstract}

  

%

\maketitle

\section{Introduction}


As an evolving infrastructure, data centers cater to multiple tenants, concurrently running diverse applications ~\cite{hpcdatacenter} like web search ~\cite{homa}, Hadoop ~\cite{hadoop}, and machine learning ~\cite{byteps}. The workloads of these applications comprise both throughput-sensitive large flows and latency-sensitive small flows simultaneously ~\cite{justitia, acc}, demanding ultra-high throughput and ultra-low latency, respectively. The substantial requirements imposed by these applications exert significant pressure on the data center networks to meet these diverse needs.



Consequently, RoCEv2 (RDMA over Ethernet version 2) has gained widespread adoption among data center designers ~\cite{rdmascale} due to its kernel bypass and hardware offloading features, significantly improving network performance compared to traditional TCP/IP. However, network designers still face numerous challenges, with the Congestion Control (CC) mechanism being a notable concern ~\cite{bolt, bfc, hpcc, dcqcn, timely, swift}. 

Especially over the past decades, the link speed has grown rapidly from 10, 50, 100, 200, and up to 400Gb/s ~\cite{ethernet}. Once congestion happens, if the control mechanism cannot respond timely, the offending traffic will aggressively occupy the limited buffer resources \cite{bfc,stephens2014practical,guo2016rdma} and seriously affects the performance of latency-sensitive flows. Moreover, priority-based Flow Control (PFC) ~\cite{pfc} is commonly used to prevent link-level packet loss when congestion control fails. PFC allows ethernet switches to avoid buffer overflows by forcing the direct upstream switch or host to pause data transmission. However, the pauses can trigger PFC deadlocks and PFC storms ~\cite{guo2016rdma, ITSY}, causing partial network entity paralysis. 

Therefore, a fast congestion control mechanism is urgently needed to reduce the network queue length and avoid PFC triggering. An ideally fast congestion control solution includes \textit{\textbf{accurate congestion detection, fast congestion notification, and accurate rate adjustment}}.

HPCC (High Precision Congestion Control) ~\cite{hpcc} is now a state-of-the-art congestion control scheme in RoCEv2 networks compared with DCQCN ~\cite{dcqcn}, Timely ~\cite{timely}, Swift ~\cite{swift}, etc. The key idea behind HPCC is to leverage the precise link load information from the in-network telemetry (INT) information to compute an accurate flow rate. HPCC senders can quickly ramp up flow rates for high utilization or ramp down flow rates for congestion avoidance. Moreover, HPCC utilizes INT information to compute the actual in-flight data volume, enabling accurate congestion detection.

Unfortunately, HPCC is sluggish in notifying the sender when congestion happens. The INT information cannot be sent directly to the sender when congestion occurs in the intermediate switch. HPCC can only send the data packet with INT to the receiver first, and then the ACK packet passes INT back to the sender. It takes the sender at least one round-trip time (RTT) to slow down when congestion occurs and another RTT to resume acceleration when congestion subsides.

\begin{figure*}[htbp]
    \centering 
    \includegraphics[width=0.98\textwidth]{"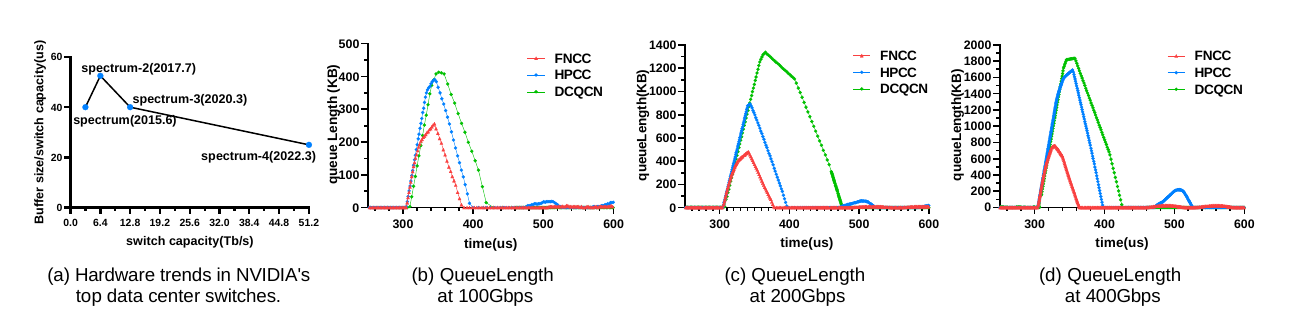"}
    \vspace{-3.0ex}
    \caption{(a), Hardware trends in NVIDIA's top data center switches. Switch capacity and link speeds have grown rapidly, but buffer sizes can not keep up with switch capacity growth. (b)$\sim$(d), Deeper queue lengths are observed across different link rates when applying HPCC and DCQCN than FNCC.}
    \label{queuelen}
\end{figure*}


\begin{figure}[htbp]
    \vspace{-1.0ex}
    \centerline{\includegraphics[width=0.50\textwidth]{"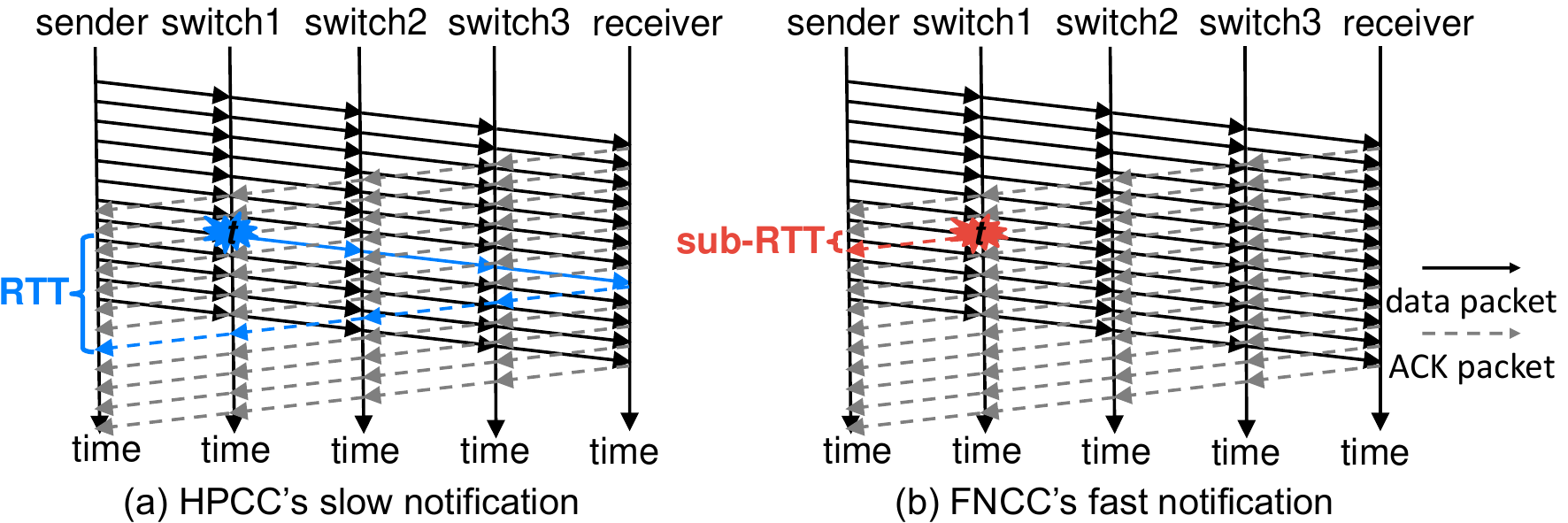"}}
    \vspace{-1.0ex}
    \caption{Notification scheme of HPCC and FNCC. Assuming congestion arises at switch1 at time t, the HPCC’s congestion information is not directly relayed to the sender.}
    \vspace{-3.0ex}
    \label{RTTsubRTT}
\end{figure}

Building upon the aforementioned observations, this paper introduces a Fast Notification Congestion Control (FNCC) mechanism, an extension of HPCC, tailored for swiftly expanding data center networks.
FNCC uses the ACK packet of the return path to carry the INT information of the request path, providing more timely and accurate INT for the sender than HPCC. The term ``request path'' refers to the route for the application's data packet transmission, while the term ``return path'' pertains to the route for ACK packet transmission. Additionally, to further accelerate the speed of last-hop congestion control, the primary source of congestion \cite{singh2015jupiter}, we propose that the receiver informs the sender of the number of concurrent congested flows. This information empowers the sender to calculate the target rate directly, facilitating the fast adjustment of flow rates to converge at the desired rate, bypassing a step-by-step approach.



Our contributions are summarized as follows:
\begin{itemize}
\item We uncover and delve into the delayed notification issue in existing congestion control mechanisms. 
\item We introduce the design of FNCC, a fast notification congestion control mechanism, and provide a thorough analysis of its design principles and detailed algorithm.
\item We assess FNCC through simulations, comparing it with DCQCN and HPCC. FNCC demonstrates advantages such as lower queue size, fewer pause frames, higher utilization, and improved fairness. Moreover, in comparison to DCQCN and HPCC, FNCC reduces Flow Completion Time (FCT) for actual data center workloads.
\end{itemize}


\section{MOTIVATION}
\subsection{Rapid growing of link speed}
The 2023 anniversary edition of the Ethernet Alliance Ethernet Roadmap \cite{ethernet} shows that the link speed has increased rapidly over the past decades. The link speed has quickly risen to 400 Gbps. 800Gbps and 1.6Tbps link speeds are now being developed. 

\begin{figure}[htbp]
\vspace{-2.0ex}   
\centerline{\includegraphics[width=0.35\textwidth]{"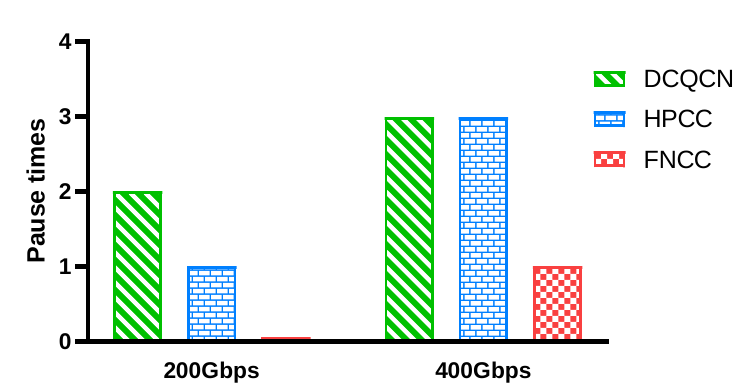"}}
\vspace{-2.0ex}   
    \caption{The count of pause frames at the congestion point. The results indicate a higher number of pause frames generated by HPCC and DCQCN algorithms than FNCC at both 200Gbps and 400Gbps rates.}
    \label{pause}
    \vspace{-3.0ex}   
\end{figure}

Nevertheless, a critical challenge persists in this escalating era of faster link speeds. Figure \ref{queuelen}a reveals a notable discrepancy between the switch buffer size and its corresponding capacity, as observed in the hardware trends of NVIDIA's top data center switches \cite{sw}. The diminishing ratio points to a concern: the switch's buffer scaling is not proportionately keeping pace with its increasing capacity. This mismatch poses a potential hurdle as it restricts the time in which the switch can absorb bursts of traffic, thereby amplifying the difficulty in managing end-to-end congestion control as these buffers will be quickly filled.

\subsection{Longer queuing delay}
\label{subsection:queuedelay}
Congestion control algorithms like DCQCN \cite{nvidianic,sw} and HPCC \cite{hpcc} have been extensively researched and deployed in data centers. Although both congestion control algorithms have been proven to improve network congestion significantly, they still face the issue of sluggish response to congestion.



\begin{figure*}[htbp]
    \centering 
    \includegraphics[width=0.9\textwidth]{"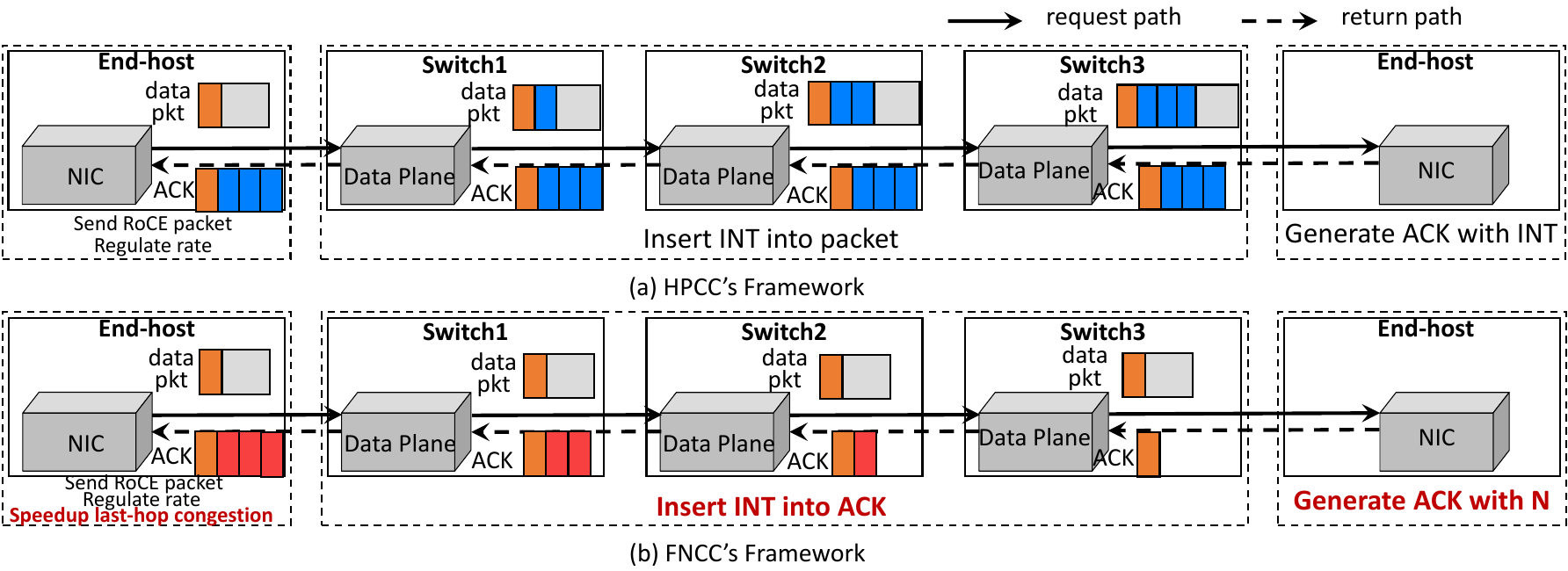"}
    \vspace{-2.0ex}
    \caption{The framework of HPCC and FNCC design. HPCC needs to add In-Network Telemetry (INT) after each data packet. The target end-host generates ACK containing all INTs and sends them back to the sender. In FNCC, the switch only adds INT to the ACK. The ACK can reach the sender faster and can provide more timely INT information. To further speed up the last-hop congestion control, FNCC’s receiver notifies the sender of the number of concurrent congested flows (N). FNCC’s sender can quickly regulate the flow rate to a fair rate.}
    \label{Figureframework}
    \vspace{-1.0ex}
\end{figure*}

As depicted in Figure \ref{RTTsubRTT}a, using HPCC as a reference, the mechanism of DCQCN closely resembles that of HPCC. If congestion arises at switch1 at time t, the congestion notification is not directly relayed to the sender. Instead, it persists in being forwarded to the receiver. Subsequently, the receiver generates an ACK packet and sends it back. This entire process is time-consuming. Any delayed response to congestion could lead to an escalation in congestion levels, resulting in a progressively deepening queue length. Figure \ref{queuelen}b$\sim$d illustrates the fluctuations in queue length at the congestion point. In this test scenario, two elephant flows are transmitted to the same switch port at the line rate. Results show that the slow response of HPCC and DCQCN resulted in deeper queue lengths across different link rates. Refer to Section \ref{subsection:micro} for a detailed experimental environment setup.

\subsection{Higher risk to trigger PFC pauses}

PFC (Priority-based Flow Control) is the most widely used flow control method in data centers to avoid packet loss. In instances where congestion arises within the downstream device's lossless queue, it promptly sends a pause frame to the upstream device to halt the flow. However, it is preferable to minimize the number of pause frames triggered by PFC, as they can affect flow fairness \cite{dcqcn}, lead to broadcast storms, and result in deadlocks \cite{rdmascale}.

Nevertheless, the long queues caused by slow congestion control will trigger PFC more frequently. Figure \ref{pause} illustrates the count of pause frames at the congestion point. The results indicate a higher number of pause frames generated by HPCC and DCQCN algorithms at both 200Gbps and 400Gbps rates. The experimental environment setup is the same as Section \ref{subsection:queuedelay}.

\subsection{Most flows are short}


Studies by \cite{homa, bfc} analyze the flow size distribution observed in the Facebook Hadoop cluster and webSearch workload. As the link speed increases, a growing percentage of traffic is swiftly processed relative to the Round-Trip Time (RTT). Moreover, it is projected that within the coming decade, a significant portion of Facebook Hadoop traffic might be completed within a single round trip.

These short-duration flows often evade control by RTT-based congestion mechanisms due to their smaller size compared to the Bandwidth Delay Product (BDP). Additionally, these flows are highly sensitive to queuing performance. Even a single erroneous or delayed congestion control decision can result in tens of microseconds of tail queuing.

\subsection{Summary}
Based on the above discussion and analysis, it becomes evident that data centers are in dire need of a fast congestion control scheme. Such a scheme is imperative to curtail the rapid filling of buffers, mitigate longer queue lengths, reduce the occurrence of pause frames, and ensure optimal performance for latency-sensitive flows.

\section{FNCC DESIGN}  

In this section, we propose a Fast Notification Congestion Control (FNCC) mechanism, which builds upon the foundation of HPCC. While HPCC boasts accurate congestion detection and rate regulation, it suffers from sluggish notification. FNCC addresses this issue by incorporating the following core designs:


\begin{itemize}
\item FNCC leverages the ACK packet of the return path to convey the INT information of the request path, which offers the sender more timely and precise INT information.  

\item The concurrent congested flows are informed to the sender so that it can quickly adjust the congested flow rate to a fair one when last-hop congestion occurs.
\end{itemize}


Concretely, Figure \ref{Figureframework} illustrates the framework of FNCC. Unlike HPCC, in FNCC, the switch is only responsible for inserting INT information into the ACK packet. To further speed up the last-hop congestion control, FNCC’s receiver notifies the sender of the number of concurrent congested flows. FNCC’s sender can quickly regulate the flow to a fair rate. 


\begin{figure}[htbp]
    \centerline{\includegraphics[width=0.49\textwidth]{"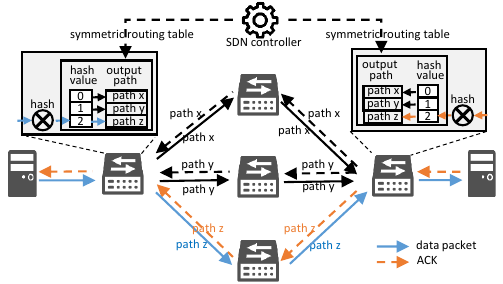"}}
    \caption{Symmetric route table. When the SDN controller builds the routing table, it should ensure that the data packet forwarding paths are in the same order as the ACK packet forwarding paths. Since the data packet and its corresponding ACK packet share the same five-tuple values, they will possess identical hash values. With a symmetric routing table in place, these packets will select the same path.}
    \label{sym}
\end{figure}

\subsection{\textbf{Design Basis}}
The proposal of the FNCC is based on the following observations.

\begin{figure}[htbp]
    \centerline{\includegraphics[width=0.23\textwidth]{"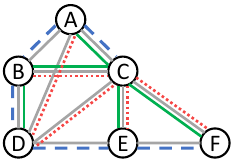"}}
    \caption{Three example spanning trees in a small network. The links are shown in gray, while the spanning trees are highlighted in color. Each spanning tree exhibits unique paths between any nodes in the network. }
    \label{spanningTree}
\end{figure}


\textbf{\textit{Observation 1. The ACK packets, traversing the congestion point, reach the sender faster.}} 

In the scenario depicted in Figure \ref{RTTsubRTT}a, suppose switch1 experiences congestion at the moment \textit{t}, HPCC does not promptly relay the existing INT (i.e., queue information) back to the sender. Its notification process consumes almost one RTT, rendering it notably slow.

Upon observing that ACK packets navigating through the congestion point reach the sender faster (as illustrated in Figure \ref{RTTsubRTT}b), FNCC capitalizes on this observation. Suppose switch1 experiences congestion at the moment \textit{t}, FNCC directly embeds INT into ACK packets on the return path, enabling fast notification of congestion situations to the source in less than an RTT.

\textbf{\textit{Observation 2. The application's data packet and its ACK packet's forwarding path can be identical.}} 

There are various topologies with path diversity, such as Fat-Tree, Jellyfish \cite{jellyfish}, Dragonfly \cite{dragonfly}, and so on. In these topologies, packets and ACK packets may choose different paths during multipath selection. However, there are several methods to address this issue.

One method is establishing a symmetric routing table where the data packet forwarding paths are in the same order as the ACK packet forwarding paths, as illustrated in Figure \ref{sym}. The Equal-Cost Multipath (ECMP) protocol is commonly utilized for load balancing among paths with identical costs. ECMP employs a hash function that operates on the five-tuple \{source IP, destination IP, source port, destination port, protocol type\}, using operations such as addition, XOR, etc., to determine the path followed by the packet. Since the data packet and its ACK packet share the same five-tuple values, they will possess identical hash values. With a symmetric routing table in place, these packets will select the same path.

Another method is constructing multiple spanning trees, each of which has a unique path between any nodes in the network, as shown in Figure \ref{spanningTree}. This approach draws inspiration from TCP-Bolt \cite{tcpbolt}, which is designed to prevent routing paths from forming loops and causing deadlocks. In our research, we employ this method to ensure that data packet and ACK packet’s path are identical.

Additionally, there are some topologies that inherently lack path diversity, where the communication path between any two nodes is unique, such as standard butterfly topology and star topology. In these topologies, the data packet and its ACK packet's forwarding path are always the same.

{\textbf{\textit{Observation 3. Application traffic, such as RC RDMA Write, may induce congestion along the request path, while their corresponding ACK packets do not contribute to congestion along the return path.
}}}

\begin{algorithm}
\caption{CP behavior}
\begin{algorithmic}[1]
\State // Port's input engine
\If{Received ACK}
    \State $ack.input\_port\_num \gets current\ port\ num;$
\EndIf
\State Forward to Ingress pipeline;

\State // Port's output engine
\If{Received ACK}
    \State $INT\_Info \gets All\_INT\_Table[ack.input\_port\_num];$
    \State Insert $INT\_Info$ to ACK packet;
\EndIf
\State Forward to Egress pipeline;
\end{algorithmic}
\end{algorithm}

Application traffic, like reliable connected RDMA Write, often carries a substantial payload, potentially reaching up to two gigabytes, rendering it particularly susceptible to network congestion. However, their ACK packets are small, containing only a few dozen bytes of data, which are used to ensure that application traffic is received reliably and do not serve to transmit application data. Therefore, the ACK packet does not cause network congestion in the opposite direction. As a result, congestion control is mainly for the congestion in the direction of application data packets transmission. 

Building upon this observation, FNCC is designed to insert the INT of the output port of the application data packets. For instance, if the data packets arrive at the switch with input port i and output port j, then its ACK's input port is j and output port is i, FNCC will insert INT of the output queue of switch port j into the ACK packet.

{\textbf{\textit{Observation 4. When last-hop congestion occurs, all flows eventually converge to a rate determined by dividing the bandwidth of the congested port by the number of flows. Moreover, it is easy for the receiver to know the number of concurrent flows.}}} 

Equation \ref{eq0} illustrates the variation of queue length \textit{q(t)} over time at the bottleneck, where \textit{W$_i$(t)} represents the sending window size of flow \textit{i}, \textit{RTT} denotes the round-trip time, \textit{W$_i$(t)/RTT} represents the input rate of flow \textit{i}, and \textit{Bandwidth} stands for the output bandwidth at the bottleneck. As the bottleneck eventually stabilizes, Equation \ref{eq0} reaches zero. At this point, the sum of the input rates of all flows equals the output bandwidth, and all flows achieve equal rates, as illustrated in Equation \ref{eq3}. 


\begin{equation}
\frac{\mathrm{d} q(t)}{\mathrm{d} t} = \sum_{i}^{N} \frac{W_i(t)}{RTT} - Bandwidth \label{eq0}
\end{equation}

\begin{equation}
   \sum_{i}^{N}\frac{W_i(t)}{RTT} = Bandwidth \label{eq3}
\end{equation}


\begin{equation}
   W_i(t) = \frac{Bandwidth*RTT}{N}  \label{eq5}
\end{equation}

This insight encourages us to regulate the sending window size \textit{W$_i$(t)} more precisely at the sender, provided we are aware of the number of concurrent flows \textit{N}, as illustrated in Equation \ref{eq5}. Fortunately, the transport layer at the receiver possesses information about the number of concurrencies, i.e., RDMA QP connections.

\subsection{\textbf{Algorithm}}
The FNCC algorithm comprises three main components: the switch (Congestion Point, CP), the sender (Reaction Point, RP), and the receiver (ACK Generation Point). In the following sections, we will delve into each of them in detail.

\subsubsection{\textbf{CP Algorithm at switch}}



HPCC relies on data packets and ACK to share information, where INT information is inserted into data packets and then forwarded by ACKs to the sender. In FNCC, it is inserted only in the ACKs. In particular, the ACK carries INT information for the request path. 




Algorithm 1 outlines the behavior of the switch. When the input engine of the switch port receives an ACK packet, it records the \textit{ack.input\_port\_number} as the current port number (Line 3) and forwards it to the ingress pipeline (Line 5). If the incoming packet is a regular data packet, it is directly forwarded to the ingress pipeline. The ingress pipeline then executes operations such as protocol checking, traffic classification, and routing.

As the ACK packet reaches the output engine, the All\_INT\_Table is looked up with the \textit{ack.input\_port\_number} as the index. \textit{$ALL\_INT\_Table$} encapsulates the INT information for all ports, and INT information contains the exact specific details as in HPCC, refer to Figure \ref{ack}. This process retrieves the \textit{INT\_Info} value for the request path (Line 8), which is then inserted into the ACK packet (Line 9) before forwarding it to the egress pipeline. In the case of a regular data packet, it is directly forwarded to the egress pipeline (Line 11). Subsequently, the egress pipeline performs operations such as flow control and quality of service (QoS) scheduling.

Note that there is no mention of multiple priority queues per port for clarity of description.

\subsubsection{\textbf{RP Algorithm at sender}}
\label{subsubsection:rp}

The FNCC's Reaction Point (RP) algorithm mirrors HPCC, employing a window-based scheme to regulate the number of in-flight bytes. Similarly, FNCC's congestion detection is also grounded in monitoring the in-flight bytes. The subsequent sections will offer a comprehensive overview of FNCC's congestion detection, rate regulation, and approach to addressing last-hop congestion scenarios.

\textbf{Congestion detection:}
\begin{equation}
     U_j = \frac{I_j}{Bandwidth_j * RTT}  \label{eq}
\end{equation}

The congestion coefficient, \textit{$U_j$}, indicates the level of congestion. When \textit{$U_j$} exceeds a threshold value, \textit{$\eta$} (where \textit{$\eta$} is nearly one, e.g., 0.95), it signifies that link \textit{j} is congested. The calculation of $U_j$ is based on Equation \ref{eq}, where \textit{$I_j$} represents all in-flight bytes for link \textit{j} and is the sum of queue length and bytes sent between two ACKs,  \textit{$Bandwidth_j$} is the bandwidth of path \textit{j}, and \textit{RTT} is the round-trip time between the sender and receiver.

\textbf{Rate regulation:}

Both HPCC and FNCC perform rate adjustments based on a window \textit{W}. As depicted in Equation \ref{eq1}, the variation in \textit{W} is contingent on the congestion coefficient \textit{$U_j$} of the most congested link. \textit{$\eta$} is a constant close to 1, e.g., 95\%, and \textit{$W_{AI}$} represents an additive increase (AI) component designed to ensure fairness, which is kept very small.

\begin{equation}
     W = \frac{W}{max_j(U_j)/\eta} + W_{AI} \label{eq1}
\end{equation}

\begin{algorithm}
\caption{RP's last-hop congestion speedup behavior }
\begin{algorithmic}[1]
    \State $U_{\text{max}} \gets 0;$
    \State $\text{hop} \gets 0;$
\Function{\text{Hop\_Detection( )}}{}
    \For{$j = 0$ to $nHop$}
        \If{$U_j > U_{\text{max}}$}
             \State $U_{\text{max}} \gets U_j;$
             \State $\text{hop} \gets j;$
             \algnotext{EndIf}
        \EndIf
        \algnotext{EndFor}
    \EndFor
\EndFunction

\Function{UpdateDATEWc}{ack}
    \State \Call{Hop\_Detection}{();}
    \If{$\text{hop} = \text{last hop} \ \& \ U_{\text{max}} > \alpha$}
        \State $W^c \gets \frac{{\text{B}} \times \text{RTT} \times \beta}{\text{ack.N}};$
    \EndIf
    \State $U_{\text{max}} \gets 0;$
    \State $\text{hop} \gets 0;$
\EndFunction

\end{algorithmic}
\end{algorithm}

To achieve a fast reaction without overreaction, both HPCC and FNCC integrate the per-ACK and per-RTT strategies. The key idea is to introduce a reference window size \textit{$W^c$}, a runtime state updated per RTT. The sender cautiously updates its window size solely upon receiving the ACK of the first packet sent with the current $W^c$. At this juncture, the sender updates \textit{$W^c$} to the current window size \textit{$W$}. Within the round-trip time, the sender updates the window size based on per-ACK information to adjust \textit{W}, see Equation \ref{eq2}.

\begin{equation}
     W = \frac{W^c}{max_j(U_j)/\eta} + W_{AI} \label{eq2}
\end{equation}

 \textbf{Fast response to last-hop congestion:}

 Inspired by our previously mentioned Observation 4, we propose a \textbf{l}ast-\textbf{h}op \textbf{c}ongestion \textbf{s}peedup (LHCS) algorithm. When congestion is detected in the last hop, the sending window of these flows is directly set to the final convergence value, offering a more agile response to the last congestion.

 In Algorithm 2, Lines 1 to 7 are dedicated to determining the location of congestion occurrence and assessing the congestion level $U_{max}$. Congestion is identified when \textit{$U_{max}$} is greater than $\alpha$ (Line 11). To prevent excessive sensitivity to the network state, the value of $\alpha$ is set slightly larger than one (e.g., 1.05). In the case of congestion occurring in the last hop, the final convergence value is assigned to $W^c$ (Line 12), where \textit{B} represents last-hop bandwidth, the value of $\beta$ is slightly smaller than one (e.g., 0.9) for draining the congested queue, \textit{N} is the number of concurrent flows. Note that the \textit{N} value is carried in the ACK packet. We will discuss it next.

\subsubsection{\textbf{ACK Generation at receiver}}

\begin{figure}[htbp]
    \centerline{\includegraphics[width=0.50\textwidth]{"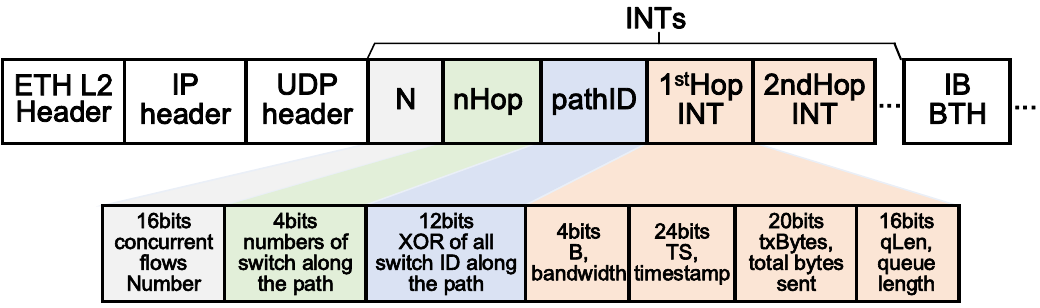"}}
    \caption{The ACK format of FNCC.}
    \label{ack}
\end{figure}

The ACK is generated by the receiver in FNCC. To help the sender converge to the target rate as soon as possible, the receiver directly informs the sender of the number of concurrent congested flows, specifically by writing the number of RDMA QP connections into the ACK packet. We define \textit{N} as the number of concurrent flows occupying 16 bits and supporting a maximum of 64k connections, which is sufficient to meet the requirements of the data center \cite{yu2011profiling}. As the ACK traverses the switch, the switch inserts INT information after the packet header of the ACK, as depicted in Figure \ref{Figureframework}b. The complete ACK format is detailed in Figure \ref{ack}.

It is worth noting that our design supports a cumulative ACK scheme (one cumulative ACK for every m consecutively received packets) at the receiver.

\begin{figure}[htbp]
    \centerline{\includegraphics[width=0.45\textwidth]{"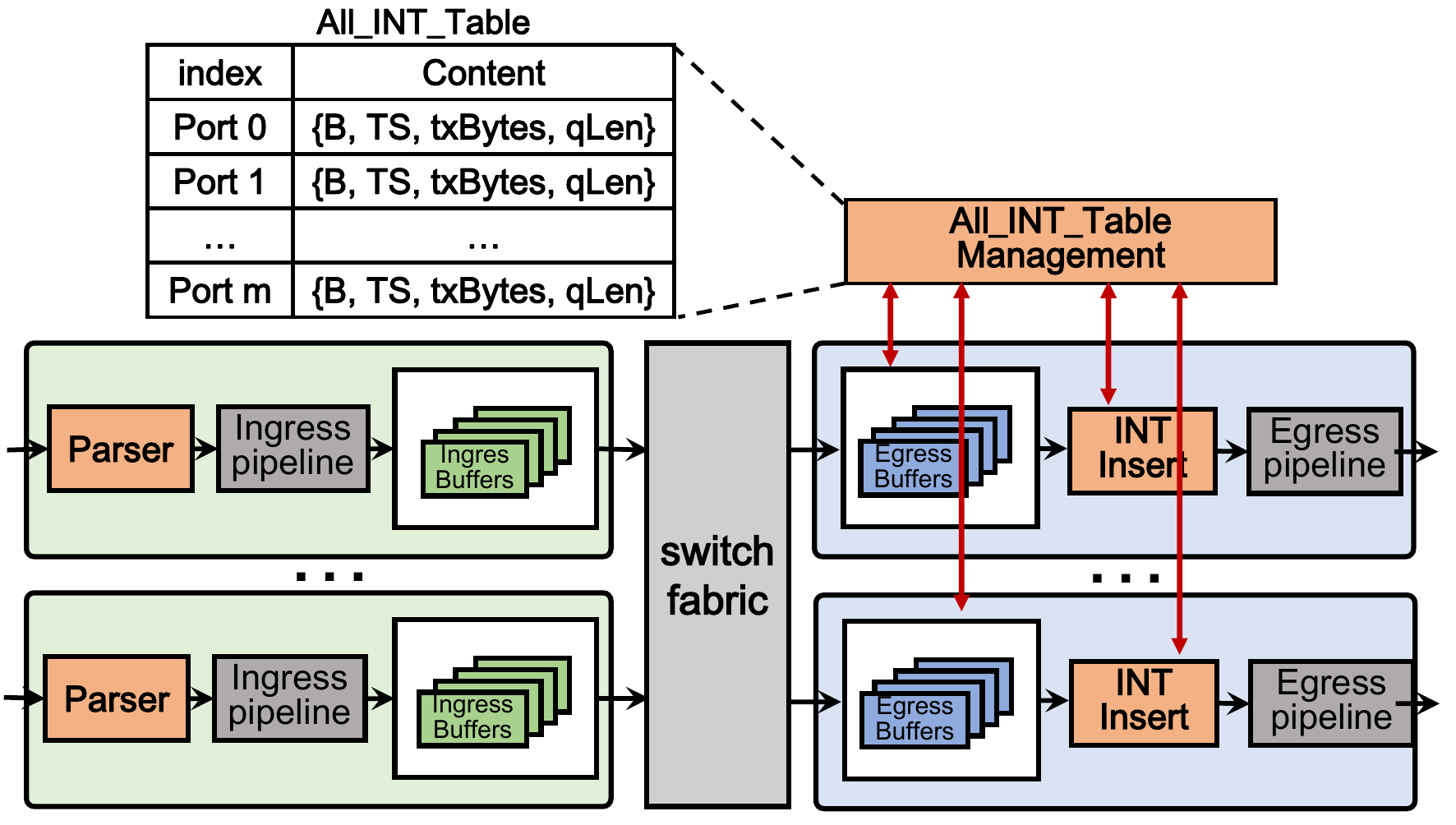"}}
    \caption{FNCC switch architecture.}
    \label{FNCC switch}
    \vspace{-3.0ex}
\end{figure}

\section{IMPLEMENTATION} 
In this section, we explore the feasibility of deploying the CP algorithm at the switch while implementing the RP algorithm and ACK generation at the host.

\subsection{CP algorithm at switch}
We give the FNCC switch architecture, as shown in Figure \ref{FNCC switch}. The key components of the CP implementation include:

\textbf{Parser module} recognizes regular data traffic and ACK packets. Once an ACK packet is recognized, record the input port number into the metadata of the ACK packet. As mentioned above, the input port number of the ACK in the return path is the output port number of application traffic in the request path.

\textbf{All\_INT\_Table Management module} is a critical system component tasked with maintaining the INT information of all output ports throughout the switch. To allow efficient lookup of the INT associated with an ACK packet, the INT is stored in a table, which is an array indexed by port number, ranging from 0 to the number of ports on the switch minus 1. This module will update \textit{All\_INT\_Table} periodically.

\textbf{INT\_Insert module} is responsible for recognizing regular traffic and ACK packets and inserting INT information into the ACK packet. The INT information is obtained by using the metadata \textit{$ack.input\_port\_num$} to look up the $All\_INT\_Table$ table. Subsequently, the processed ACK is forwarded to the Egress pipeline.


 \begin{figure*}[htbp]
    \centering 
    \includegraphics[width=0.95\textwidth]{"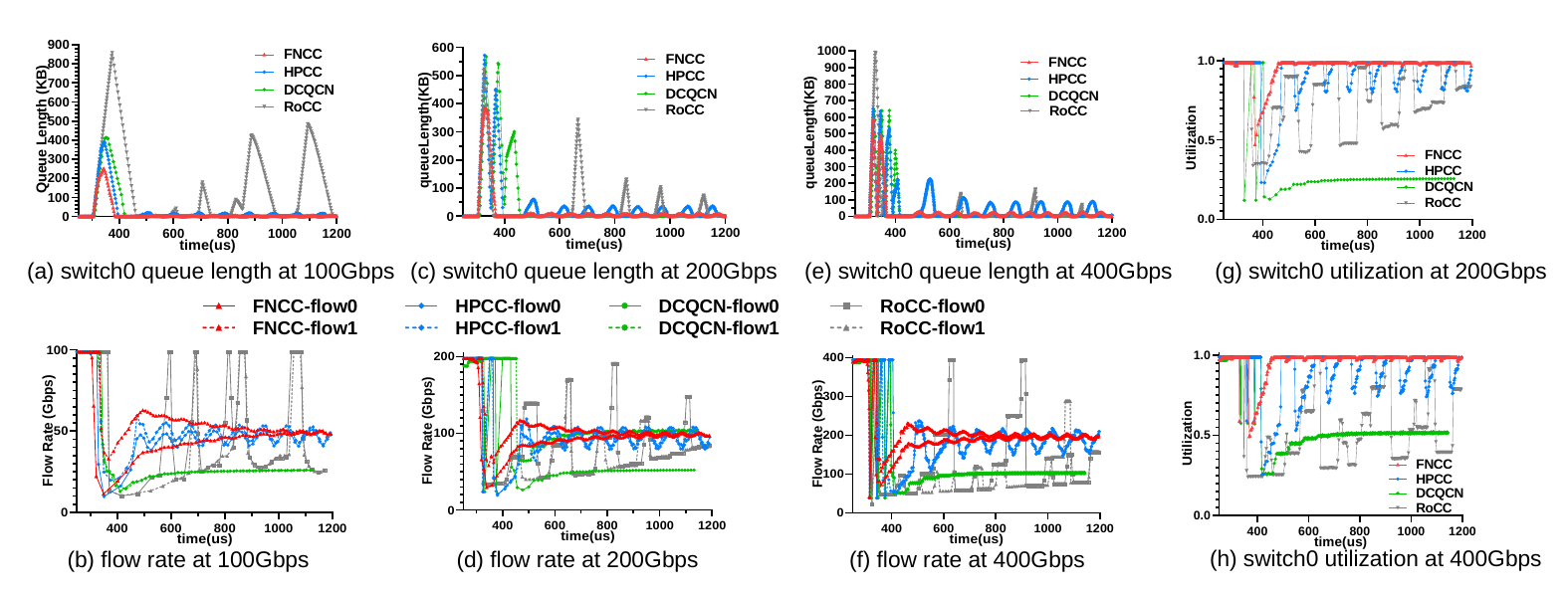"}
    \vspace{-2.0ex}
    \caption{(a)$\sim$(b), Queue length's variation at congestion point and flow rate's variation at senders at 100 Gbps. FNCC is the first to detect congestion and slow down. The queue length at FNCC's congestion point is the lowest, and the two elephant flows of FNCC are the first to converge to a fair rate. (c)$\sim$(f), Queue length's variation at congestion point and flow rate's variation at senders at 200Gbps and 400Gbps. FNCC is also the fastest to react. (g)$\sim$(h) FNCC can maintain the highest utilization.}
    \label{flowratequeuesize124002}
\end{figure*}

\subsection{RP algorithm and ACK generation at host}
Compared to HPCC, FNCC's RP algorithm at the sender host incorporates congestion hop detection and $W^c$ update functions, with low computational costs and without requiring additional buffer resources. Meanwhile, at the receiver host, FNCC only inserts the number of concurrent flows, \textit{N}, into the ACK packet, also incurring minimal computational overhead.

\subsection{Hardware feasibility}
FNCC requires a memory size proportional to the number of physical output queues in the switch to store the \textit{All\_INT\_Table} table. Each entry in the table is 64 bits width \{4 bits B, 24 bits TS, 20 bits txBytes, 16 bits qLen\}. Assuming a 64-port switch, the size of the INT table is 4k bits, which is small compared to the switch's shared buffer size (3.2MB, 6.4MB, 12.8MB, 51.2MB, etc.).

Implementing FNCC in switches is both simple and cost-effective. It only requires additional resources to implement its module's design, which is abundantly available in today's commodity switches \cite{p4fpga, NetFPGA, Trident3}. The process of looking up the INT table and inserting INT into ACK requires only \(\mathcal{O}\)(1) time. With an increasing number of switches supporting programmable and open data planes, we anticipate that the \textit{All\_INT\_Table} table lookup and insertion procedures can be executed at the data plane at a line rate.

FNCC's implementation in hosts is also straightforward and cost-effective. At the sender host, FNCC incorporates congestion hop detection and $W^c$ update functions. At the receiver host, FNCC only inserts the number of concurrent flows. Those introduce minimal computational and storage overhead.

\section{PERFORMANCE EVALUATION} 

In this section, we evaluate the performance of FNCC through the INET model suite on the OMNeT++ \cite{omnet} simulator. We use different traffic patterns and topologies to assess the effectiveness of FNCC, utilizing traffic workloads derived from publicly available data center traffic traces.

\begin{figure}[htbp]
    \vspace{-2.0ex}
    \centerline{\includegraphics[width=0.45\textwidth]{"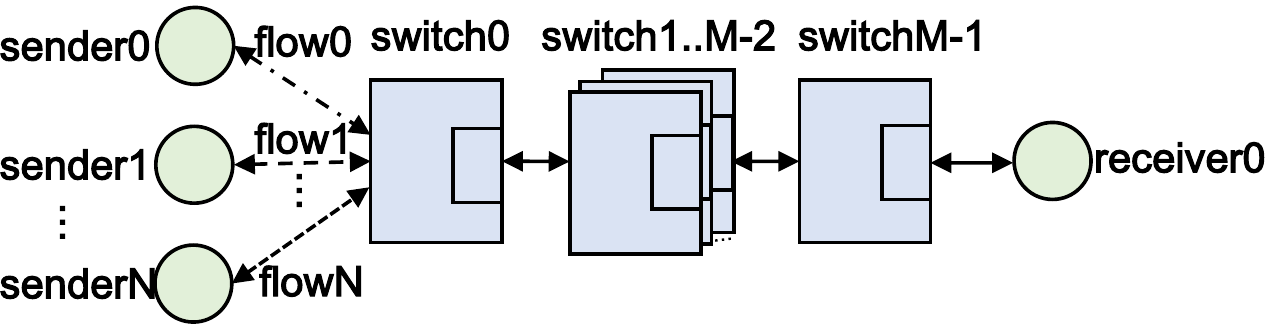"}}
    \caption{Typical dumbbell topology.}
    \label{typical_topo}
    \vspace{-2.0ex}
\end{figure}

All links in all topologies are running at the 100Gbps rate with 1.5$\mu$s propagation delay. To eliminate the impact of RoCEv2 QoS, packets from all sources are transferred on the same service level (SL) or virtual lane (VL). The Maximum Transmission Unit (MTU) is set to 1518 bytes. We implement several state-of-the-art solutions, including DCQCN, HPCC, and RoCC. DCQCN has been implemented in Mellanox commercial network adapters and switches \cite{nvidianic,sw}. HPCC is already employed in Alibaba's predictable network \cite{hpcc}. RoCC is a typical switch-driven congestion control proposed by Cisco. At the switch, RoCC uses the proportional-integral (PI) controller to compute the fair rate. DCQCN and RoCC's parameters are assigned to the default values recommended in research \cite{dcqcn, rocc}.

We conduct five types of experiments to evaluate FNCC: 
\begin{itemize}
\item Response speed on micro-benchmarks
\item Robustness against higher line rates
\item Fairness over multiple flows
\item Gains across various congestion scenarios
\item Large-scale simulation resembling a real data center


\end{itemize}

\subsection{Response speed on micro-benchmarks}
\label{subsection:micro}



\textbf{Typical scenario.} As shown in Figure \ref{typical_topo}, we build a typical dumbbell topology network. The experiment includes three switches (M = 3) and two senders (N = 2). The first elephant flow, flow0, starts at time 0 at 100Gbps. And a new elephant flow, flow1, joins at 300us at 100Gbps. To prevent packet loss, we enable PFC and set the PFC threshold to 500KB. We monitor the rate changes of the two elephant flows at the sender to visualize the congestion response speed. Moreover, we also observe the queue length of the congestion port of switch1.



\textbf{FNCC has a faster response to congestion}. Figure \ref{flowratequeuesize124002}b
illustrates the behavior of RoCC, DCQCN, HPCC, and FNCC. FNCC is the first to slow down at 300$\mu$s, followed by HPCC at 330$\mu$s, DCQCN at 346$\mu$s, and RoCC at 370$\mu$s. It is demonstrated that our scheme of utilizing the ACK packet carrying INT information is not only theoretically faster but also empirically.

\textbf{FNCC has faster convergence to fair rates.} When congestion is relieved at the switch, from Figure \ref{flowratequeuesize124002}b, the two elephant flows applying FNCC have faster and smoother convergence to fair rates. However, the two flows applying HPCC converge to a fair rate jitterily. When using DCQCN, the two flows are slow to recover. And RoCC is hard to converge at the microsecond level.

\textbf{FNCC has the shallowest queue depth.} From Figure \ref{flowratequeuesize124002}a, due to the advantage of fast notification in FNCC, it can quickly reduce the sender's rate and minimize the queue length peak at congestion points.



\subsection{Robustness against higher line rates}
\label{subsection:robust}

One of the goals of the FNCC is to remain robust in the face of increasing data center line rates. We repeat the simulation in Section \ref{subsection:micro}, increasing the link capacity from 100Gbps to 200Gbps and 400Gbps. It is difficult for the switch to maintain a small queue.


\textbf{Low queue length.}
As shown in Figure \ref{flowratequeuesize124002}c and Figure \ref{flowratequeuesize124002}e, FNCC always maintains the shortest queue length at the congestion point compared with RoCC, DCQCN and HPCC. FNCC is always the first to slow down and converge to a fair rate, as shown in Figure \ref{flowratequeuesize124002}d, \ref{flowratequeuesize124002}f. This proves that the fast notification scheme of FNCC is robust even at high rates.

\textbf{High utilization.} 
From Figure \ref{flowratequeuesize124002}g and Figure \ref{flowratequeuesize124002}h, FNCC consistently maintains the highest utilization. Once congestion is relieved, FNCC promptly updates senders with the latest queue length at the congestion point, enabling them to promptly increase speed to avoid underutilized links.


\textbf{Few pause frame.} 
We measured the number of pause frames generated at the congestion point using FNCC, HPCC, and DCQCN at 200Gbps and 400Gbps. From Figure \ref{pause}, it can be seen that the number of pause frames generated when applying the FNCC algorithm is the lowest at both 200Gbps and 400Gbps rates.

\subsection{Fairness over multiple flows}

\begin{figure*}[htbp]
    \centering 
    \includegraphics[width=0.95\textwidth]{"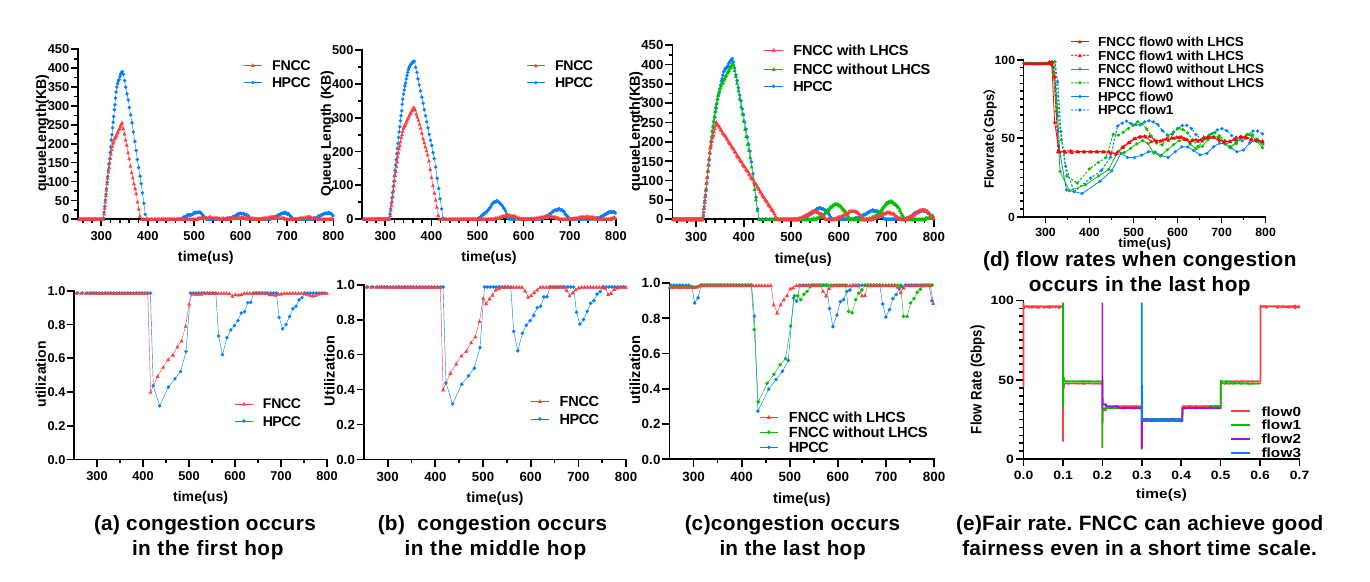"}
    \vspace{-2.0ex}
    \caption{ (a)$\sim$(c), FNCC maintains lower queue depths while guaranteeing higher link utilization. FNCC reduces queue depth by 37.5\% in the first-hop congestion, 29.5\% in the middle-hop congestion, 8.4\% in the last-hop congestion without enabling Last-Hop Congestion Speedup (LHCS), and 38.5\% in the last-hop congestion with LHCS. (d) FNCC with LHCS can sensitively detect the last-hop congestion and quickly adjust to a fair rate * $\beta$. The value of $\beta$ is slightly smaller than one (e.g., 0.9) for draining the congested queue. (e), FNCC can achieve good fairness. }
    \label{fst-mdl-last}
    \vspace{-2.0ex}
\end{figure*}

To verify the fairness of FNCC, we adopt a dumbbell topology as shown in Figure \ref{typical_topo}. The experiment includes four senders (N = 4). They send long-lived flows to receiver0. Every 100 milliseconds, a new sender initiates a new flow and subsequently exits in sequence. Figure \ref{fst-mdl-last}e demonstrates the fair throughput of these four flows. The experimental results prove that FNCC can achieve good fairness even in a short time scale.


\begin{figure}[htbp]
    \centerline{\includegraphics[width=0.30\textwidth]{"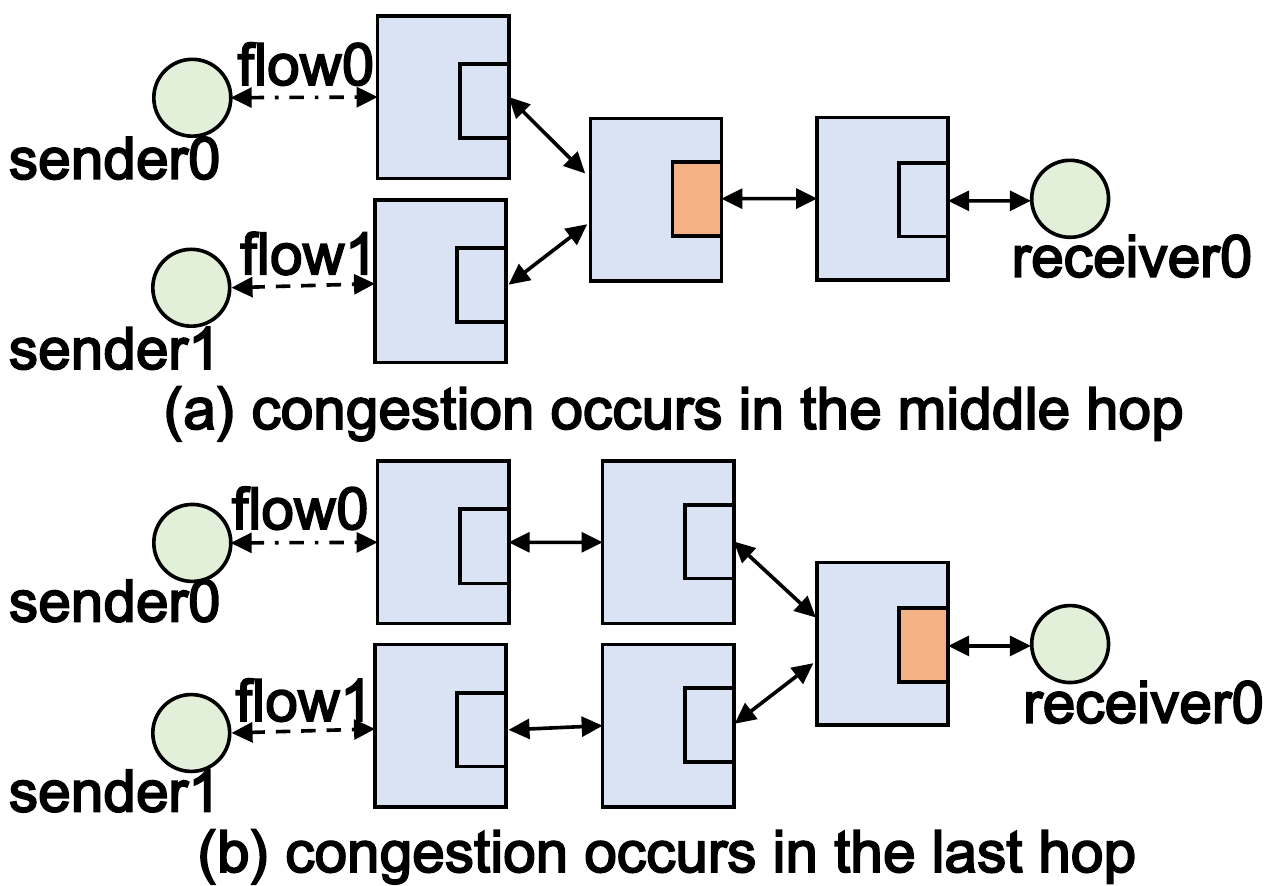"}}
    \caption{Topology: Congestion occurs in different locations. }
    \vspace{-3.5ex}
    \label{mdl-last-topology}
\end{figure}

\begin{figure}[htbp]
    \centerline{\includegraphics[width=0.42\textwidth]{"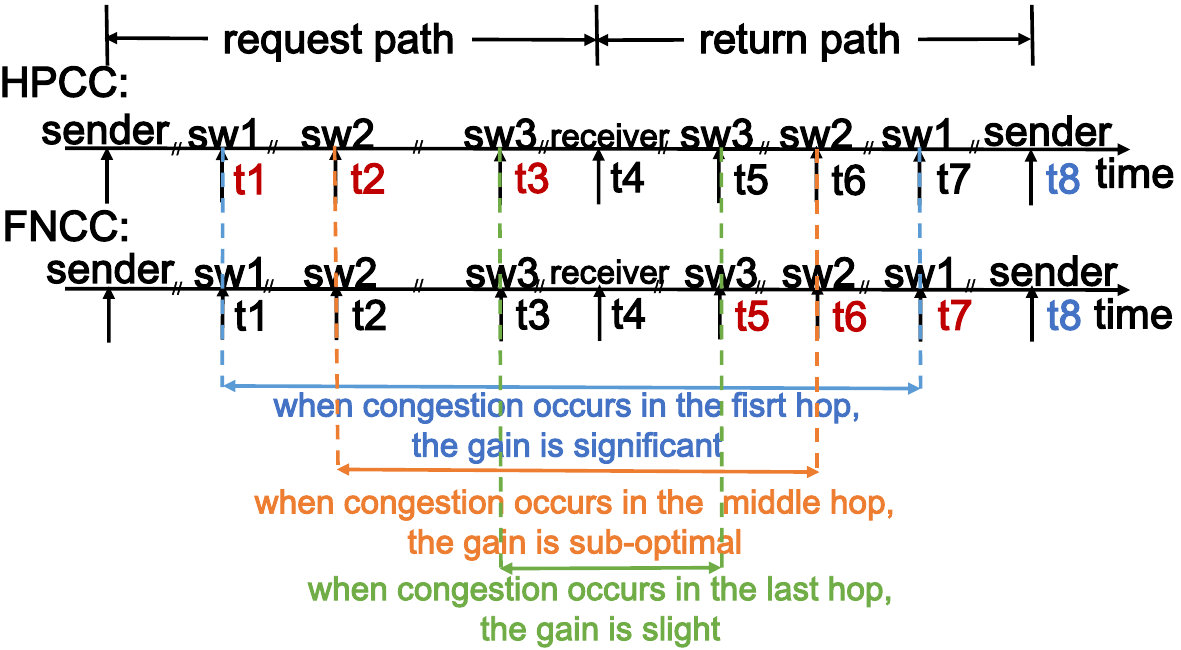"}}
    \caption{Theoretical model. }
    \vspace{-3.5ex}
    \label{timeline}
\end{figure}

\subsection{Gains across various congestion scenarios}

We extended the experiment with congestion occurring in the middle hop and the last hop, both with a link rate of 100Gbps. The experimental topologies are shown in Figure \ref{mdl-last-topology}.

\subsubsection{Theoretical Analysis}

In Figure \ref{timeline}, HPCC inserts INTs at switches along the request path at moments t1, t2, and t3. The sender receives an ACK message with INT and initiates rate adjustment at the moment t8. On the other hand, FNCC inserts INTs at the switch along the return path at moments t5, t6, and t7. The INT values inserted at t5, t6, and t7 are more timely, allowing congestion or under-utilization to be captured earlier and more promptly when the sender performs rate adjustment at t8.


However, the rate adjustment is determined by the INT of the most congested link, as discussed above. In the scenario where congestion occurs on sw1 (switch1), FNCC exhibits an even more pronounced advantage over HPCC due to the significant time difference between t7 and t1. If congestion arises on sw2 (switch2), the time difference between t6 and t2 is relatively large, resulting in FNCC having a sub-optimal advantage over HPCC. In the case of congestion on sw3 (switch3), where the time difference between t5 and t3 is not very prominent, the FNCC advantage over HPCC is marginal. However, for the last-hop congestion scenario, we propose the ``last-hop congestion speedup algorithm'' in Section \ref{subsubsection:rp}. The sender can quickly make precise rate adjustments.

\subsubsection{Experimental results}


According to Figure \ref{fst-mdl-last}, while maintaining a higher link utilization, FNCC reduces queue depth by 37.5\% in first-hop congestion, 29.5\% in middle-hop congestion, 8.4\% in last-hop congestion without enabling Last-Hop Congestion Speedup (FNCC without LHCS), and 38.5\% in last-hop congestion with enabling Last-Hop Congestion Speedup (FNCC with LHCS). We also compared the variations of flow rate when using HPCC, FNCC without LHCS, and FNCC with LHCS. The experimental results are shown in Figure \ref{fst-mdl-last}d. At 300us, the addition of flow1 causes the last-hop switch to become congested, and the FNCC with last-hop speedup can be quickly adjusted to a fair rate * $\beta$. The value of $\beta$ is slightly smaller than one (e.g., 0.9) for draining the congested queue and maintaining high utilization. After 450us, the last-hop congestion disappears, and the rate of the two flows is no longer constrained by the last-hop speedup Algorithm 2, but is controlled by Equation \ref{eq2} and gradually adjusted to a fair rate. To facilitate readability, we have omitted the comparison with DCQCN and RoCC. From the results shown in Figure \ref{flowratequeuesize124002}, it is evident that HPCC outperforms DCQCN and RoCC. 



\begin{figure*}[htbp]
    \centering
    \includegraphics[width=\textwidth]{"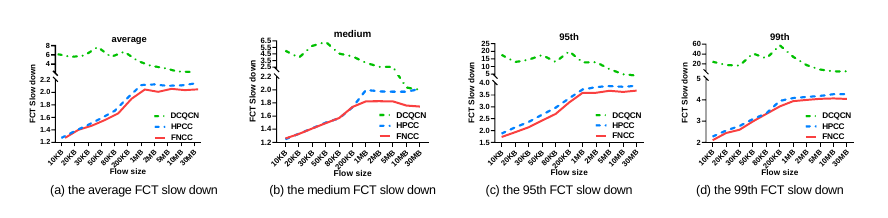"}
    \vspace{-6.0ex}
    \caption{The average, medium, 95th, and 99th FCT slowdown of DCQCN, HPCC, and FNCC with WebSearch (50\% avg. load). FNCC always has the lowest tail latency. Especially for flows larger than 1 MB, FNCC has a 12.4\% and 42.8\% reduction in medium FCT slowdown than HPCC and DCQCN, respectively.}
    \label{websearch}
\end{figure*}

\begin{figure*}[htbp]
\vspace{-5.0ex}
    \centering
    \includegraphics[width=\textwidth]{"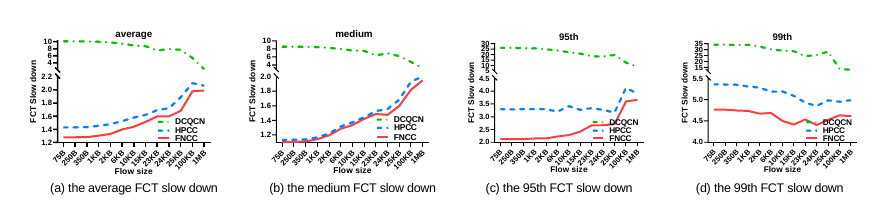"}
    \vspace{-6.0ex}
    \caption{The average, medium, 95th, and 99th FCT slowdown of DCQCN, HPCC and FNCC with Hadoop (50\% avg. load). For flows shorter than 100KB, FNCC has a 27.4\% and 88.9\% reduction in 95th FCT slowdown than HPCC and DCQCN, respectively.}
    \vspace{-2.0ex}
    \label{hadoop}
    
\end{figure*}

\subsection{Large-scale simulations} 
\label{subsection:largescale}

We use large-scale simulations to evaluate FNCC and compare it with DCQCN \cite{dcqcn} and HPCC \cite{hpcc} regarding FCT slowdown. ``FCT slowdown'' means a flow's actual FCT normalized by its ideal FCT when the network only has this flow. Due to the poor performance of RoCC at the microsecond level and the fact that it has not been used in practice, we overlooked the comparison with it. We use a three-level fat-tree topology (k = 8) with 128 servers. Each server has a single 100Gbps NIC connected to a single top-of-rack (ToR) switch. The capacity of each link between core and aggragation switches, aggragation switches and ToR switches are all 100Gbps (1:1 oversubscription). All links have a 1.5µs propagation delay. We implement ECMP on the ToR and aggragation switches to distribute the load across the links equally. We use traffic loads derived from two publicly available data center traffic distributions consisting of throughput-sensitive large flows (WebSearch traffic) and latency-sensitive small flows (FB\_Hadoop traffic) \cite{homa}. We run our simulations using 50\% average link load levels and repeat each experiment 5 times. The FCTs slowdown we present are the average values of the five sets of results.

Figure \ref{websearch} and Figure \ref{hadoop} respectively show the average, medium, 95th percentile, and 99th percentile FCT slowdown for WebSearch traffic and FB\_Hadoop traffic at 50\% average load. 


\textbf{FCT slowdown.} 
Based on the average percentile FCT, as shown in Figure \ref{websearch}a and Figure \ref{hadoop}a, FNCC outperforms DCQCN and HPCC for all the flow sizes. 

\textbf{FNCC is beneficial to long flows.} 
Our fast notification mechanism not only swiftly relieves congestion but also prevents low link utilization. Once congestion is alleviated, FNCC can quickly increase speed to avoid under-utilization of links, which is especially beneficial for long flows. As depicted in Figure \ref{websearch}a and Figure \ref{websearch}b, for flows larger than 1 MB, FNCC can reduce congestion by up to 12.4\% compared to HPCC and 42.8\% compared to DCQCN.

\textbf{FNCC is also beneficial to short flows.} Since FNCC has a fast notification mechanism and keeps the shallowest queues, it is beneficial to short flows mostly. HPCC and DCQCN schemes endure slow notification, leading to slow congestion relief, so they cannot keep the latency low. Figure \ref{hadoop}c shows that for the flows shorter than 100KB, FNCC achieves a much lower FCT slowdown than other schemes at the 95th percentile. Specifically, FNCC has a 27.4\% and 88.9\% reduction in FCT slowdown than HPCC and DCQCN, respectively.

\vspace{-2.0ex}
\section{RELATED WORK}
Congestion control is an enduring topic, and here we briefly introduce some closely related work.

\textbf{Switch-driven notification.}
RoCC \cite{rocc} and BFC \cite{bfc} are two typical mechanisms whose notification entity is located at the switch. RoCC leverages a proportional-integral controller to compute the fair rate of active flows. BFC achieves accurate per-hop per-flow flow control. However, RoCC relies on at least nine parameters to tune, and requires millisecond-level delays to converge, see Figure \ref{flowratequeuesize124002}. BFC requires large buffer resources, as it necessitates setting up separate queues for each flow, which poses a challenging implementation in switches.





\textbf{Receiver-driven notification.}
To prevent the last-hop congestion, certain proposals advocate relocating the notification entity to the receiver side. ExpressPass \cite{expresspass} avoids congestion by directly dispatching credits to senders. However, managing distinct timers on RDMA NICs to orchestrate credit pacing for each flow poses challenges. Homa \cite{homa} dynamically schedules packets using priority queues. Nevertheless, due to the likelihood of packet reordering with the packet spraying scheme, more robust support is needed in RDMA networks.

\textbf{End-to-end notification.}
End-to-end based solutions like DCQCN \cite{dcqcn}, Timely \cite{timely}, Swift \cite{swift}, and HPCC \cite{hpcc}, etc., rely on rate regulation through congestion feedback mechanisms like ECN, RTT, and INT. However, a shared drawback among these approaches is their delayed reaction to congestion. We leverage HPCC's benefits of accurate flow rate updates. To address its slow congestion notification, we introduce FNCC, offering more timely and precise INT to the sender.



\section{Conclusion}
We discuss the limits of existing congestion control methods and propose FNCC, a novel fast notification congestion control mechanism. Compared to HPCC, FNCC can provide a more timely and accurate INT for the sender by utilizing the ACK packet of the return path to carry INT information of the request path. Additionally, FNCC further accelerates the speed of last-hop congestion control, by promptly informing the sender of the number of concurrent congested flows and adjusting the congested flow to a fair rate. Experimental results show that FNCC not only achieves lower queue size, fewer pause frames, and higher utilization even at 400Gbps, but also reduces FCT for real data center workloads compared with DCQCN and HPCC.

\bibliographystyle{ACM-Reference-Format}
\bibliography{myrefs}

\newpage
\appendix
\section{The overall process of Congestion Control at the sender side.}

Algorithm 3 outlines the complete procedure of Congestion Control on the sender side for a single flow. Upon receiving each new ACK message, the procedure NewACK is triggered at Line 41. At Line 42, the variable ``lastUpdateSeq” is utilized to store the sequence number of the first packet sent with a new $W^c$. For a new synchronization between $W^c$ and W, the sequence number in the incoming ACK must be larger than ``lastUpdateSeq'' (as indicated in Line 33-34 and Line 37-38). The sender calculates a new window size, W, at Line 43 or Line 46. 

Function MeasureInFlight calculates the normalized in-flight bytes at Line 8. Initially, it calculates the transmission rate (txRate) for each link based on the current and previously accumulated transmitted bytes (txBytes) and the corresponding timestamp (ts) (Line 7). Additionally, it employs the minimum of the current and previous queue length to eliminate noise in qlen (Line 8). The loop from Lines 10 to 13 selects the maximum value of $U_i$. Instead of directly utilizing $U_i$, an Exponentially Weighted Moving Average (EWMA) is applied at Line 13 to filter out noise arising from timer inaccuracies and transient queues.

Function Hop\_Detection is dedicated to determining the location of congestion occurrence and assessing the congestion level (Line 17-20). Note that if congestion is detected in the last hop, the sending window of $W^c$ is directly set to the final convergence value by Function UpdateWc (Line 25).

Function ComputeWind, in addition to containing Function UpdateWc to speed up last-hop congestion control, also integrates multiplicative increase/decrease (MI/MD) and additive increase (AI) to strike a balance between responsiveness and fairness. When a sender determines the need to increase the window size, it initially employs AI for a maximum of maxStage attempts with a step size of $W_{AI}$ (Line 36). If there is still room for increase after maxStage AI attempts or if the normalized in-flight bytes exceed $\eta$, the window size is promptly adjusted upward or downward (Line 31-32).

\begin{algorithm}
\caption{Overall algorithm at the sender} 
\begin{algorithmic}[1]
\State $U[5] \gets 0;$
\State $U_{\text{max}} \gets 0;$
\State $\text{hop} \gets 0;$

\Function{MeasureInFlight}{ack}
    \State $u \gets 0;$
    \For{each link $i$ on the path}
        \State $\text{txRate} \gets \frac{\text{ack.L}[i].\text{txBytes} - L[i].\text{txBytes}}{\text{ack.L}[i].\text{ts} - L[i].\text{ts}};$
        \State $u' \gets \frac{\min(\text{ack.L}[i].\text{qlen}, L[i].\text{qlen})}{\text{ack.L}[i].B.T} + \frac{\text{txRate}}{\text{ack.L}[i].B};$
        \State $U_i \gets u';$
        \If{$u' > u$}
            \State $u \gets u'; \tau \gets \text{ack.L}[i].\text{ts} - L[i].\text{ts};$
            \State $\tau \gets \min(\tau, T);$
            \State $U \gets (1 - \frac{\tau}{T}) \cdot U + \frac{\tau}{T} \cdot u;$
             \algnotext{EndIf}
        \EndIf
         \algnotext{EndFor}
    \EndFor
    \State \textbf{return} $U$
\EndFunction

\Function{Hop\_Detection( )}{}
    \For{$j = 0$ to $nHop$}
        \If{$U_j > U_{\text{max}}$}
             \State $U_{\text{max}} \gets U_j;$
             \State $\text{hop} \gets j;$
        \EndIf
    \EndFor
\EndFunction

\Function{UpdateWc}{ack}
    \State \Call{Hop\_Detection}{();}
    \If{$\text{hop} = \text{last hop} \ \& \ U_{\text{max}} > \alpha$}
        \State $W^c \gets \frac{{\text{ack.L[0].B}} \times \text{RTT} \times \beta}{\text{ack.N}};$
    \EndIf
    \State $U_{\text{max}} \gets 0;$
    \State $\text{hop} \gets 0;$
\EndFunction

\Function{ComputeWind}{U, updateWc, ack}
    \State \Call{UpdateWc}{ack};
    \If{$U \geq \eta$ or $\text{incStage} \geq \text{maxStage}$}
        \State $W \gets \frac{W^c}{U/\eta} + W_{\text{AI}};$
        \If{updateWc}
            \State $\text{incStage} \gets 0; W^c \gets W;$
            \EndIf
    \Else
        \State $W \gets W^c + W_{\text{AI}};$
        \If{updateWc}
            \State $\text{incStage} \gets \text{incStage} + 1; W^c \gets W;$
        \EndIf
    \EndIf
            
    \State \textbf{return} $W$
\EndFunction

\Procedure{NewAck}{ack}
    \If{$\text{ack.seq} > \text{lastUpdateSeq}$}
        \State $W \gets \text{ComputeWind}(\text{MeasureInFlight}(ack), \text{True}, ack);$

        \State $\text{lastUpdateSeq} = \text{snd\_nxt};$
    \Else
        \State $W \gets \text{ComputeWind}(\text{MeasureInFlight}(ack), \text{False}, ack);$
    \EndIf
    \State $R \gets \frac{W}{T}; L \gets \text{ack.L}$
\EndProcedure
\end{algorithmic}
\end{algorithm}

    
\end{sloppypar}

\end{document}